\documentclass[12pt]{article}
\usepackage{graphicx,latexsym,psfrag,multirow,cite}

\newcommand{\bea}{\begin{eqnarray}}
\newcommand{\eea}{\end{eqnarray}}
\newcommand{\beq}{\begin{equation}}
\newcommand{\eeq}{\end{equation}}
\newcommand{\simgt}{\hbox{ \raise3pt\hbox to 0pt{$>$}\raise-3pt\hbox{$\sim$} }}
\newcommand{\simlt}{\hbox{ \raise3pt\hbox to 0pt{$<$}\raise-3pt\hbox{$\sim$} }}

\def\m{$\overline m$} \def\mpole{$m_{\rm pole}$}
\def\Etot{$E_{\rm tot}$}
\begin{document}
\begin{titlepage}
\title{Comparing the QCD potential in Perturbative QCD \\
and Lattice QCD at large distances \vspace{2cm}}
\author{S.~Recksiegel$^1$ and Y.~Sumino$^2$
\\
\\ $^1$ Theory Group, KEK\\
Tsukuba, Ibaraki, 305-0801 Japan\footnote{From March 2003: 
Physikdepartment T31, Technische Universit\"at M\"unchen, D--85747 Garching, Germany}\\
\\ $^2$ Department of Physics, Tohoku University\\
Sendai, 980-8578 Japan
}
\date{}
\maketitle
\thispagestyle{empty}
\vspace{-5truein}
\begin{flushright}
{\bf hep-ph/0212389}\\
{\bf TU--673}\\
{\bf KEK--TH--860}\\
{\bf December 2002}
\end{flushright}
\vspace{4.5truein}
\begin{abstract}
\noindent
{\small
We compare the perturbatively calculated QCD potential to that
obtained from lattice calculations in the theory without
light quark flavours.
We examine $E_{\rm tot}(r) = 2 m_{\rm pole} + V_{\rm QCD}(r)$
by re-expressing it in the $\overline{\rm MS}$ mass
$\overline{m} \equiv m^{\overline{\rm MS}}(m^{\overline{\rm MS}})$
and by choosing specific prescriptions for fixing the
scale $\mu$ (dependent on $r$ and \m).
By adjusting \m\ so as to maximise the range of convergence, we
show that perturbative and lattice calculations agree up to
$3\, r_0\simeq 7.5 \,{\rm GeV}^{-1}$ 
($r_0$ is the Sommer scale) within the 
uncertainty of order $\Lambda_{\rm QCD}^3\, r^2$.
}
\end{abstract}
\vfil
\end{titlepage}
  
%%%%%%%%%%%%%%%%%%%%%%%%%%%%%%%%%%%%%%%%%%%%%%%%%%%%
\section{Introduction}

For decades, the static QCD potential $V_{\rm QCD}(r)$, formally
defined from an expectation value of the Wilson loop,
has been widely studied 
for the purpose of elucidating the nature of the interaction between heavy
quark and antiquark.
In modern language, a link to physical reality can be made naturally
in the frame of potential--nonrelativistic
QCD (potential--NRQCD) formalism  \cite{pinedasoto,bpsv,nnnlo-H}, in which
$V_{\rm QCD}(r)$ is identified with the leading potential
in expansion in $1/m$ of the heavy quarkonium system.
Therefore, $V_{\rm QCD}(r)$ dictates, for instance, the bulk of the
spectra of the bottomonium and charmonium states.

%Computations of the QCD potential within perturbative QCD have
%been carried out over time \cite{}.
%Meanwhile, there have been
%a number of studies to extract the potential between heavy quarks
%from the spectra of the quarkonium states
%within phenomenological potential model frameworks
%(see e.g.\ \cite{}).
%More recently, fairly accurate calculations
%of the QCD potential 
%have been provided by lattice simulations \cite{bsw,ukcollab}.
%In the distance region
%relevant to phenomenological studies of the bottomonium and charmonium
%states,
%$0.5~{\rm GeV}^{-1} \simlt r \simlt 5~{\rm GeV}^{-1}$,
%the latter two types of studies gave mutually consistent shapes of the
%potentials.
%On the other hand, the perturbative QCD prediction of $V_{\rm QCD}(r)$
%was barely convergent and, moreover, seemingly qualitatively incompatible with
%the other potentials at
%$r \simgt 0.5~{\rm GeV}^{-1}$.

Some time ago, there was a breakthrough 
that drastically improved the predictive power
of perturbative QCD for the QCD potential and the heavy quarkonium
spectrum:
the perturbative predictions for 
these quantities became much more accurate.
This was achieved by properly eliminating 
contributions from infrared (IR) degrees of freedom in the computations
\cite{renormalon1,renormalon2}.
The central quantity is the total energy of a
static quark--antiquark pair, defined by the sum of the
quark and antiquark pole masses and the QCD potential,
$E_{\rm tot}(r) = 2 m_{\rm pole} + V_{\rm QCD}(r)$.
We can achieve the decoupling of IR degrees of freedom 
(renormalon cancellation)
at each order of the perturbative expansion by
(1) re--expressing the quark pole mass in terms of a
so--called short--distance mass, such as the $\overline{\rm MS}$ mass,
and 
(2) expanding $m_{\rm pole}$ and $V_{\rm QCD}(r)$ in the same
coupling constant.\footnote{
This is somewhat involved technically, since usually 
$m_{\rm pole}$ and $V_{\rm QCD}(r)$
are expressed in terms of different coupling constants.
}
As a result, the perturbative predictions
become stable against a variation of the renormalisation
scale $\mu$, and also the perturbative series show a much better convergence
behaviour, as compared to those in the conventional
computations.

It was then natural to compare the perturbative QCD predictions 
with existing experimental data or 
with other theoretical predictions which incorporate 
non--perturbative effects.
The main aim of this program is to clarify the
differences between the perturbative QCD predictions and the full QCD
predictions, given the more accurate predictions of the former.
The first comparison \cite{bsv1} was made for the bottomonium spectrum
(and also for part of the charmonium spectrum) between the 
perturbative prediction and the experimental data.
It was followed by a comparison \cite{Sumino:2001eh}
between the perturbative QCD prediction
of the QCD potential and typical phenomenological
potentials (used in phenomenological approaches to heavy quarkonium
physics),
and then by a comparison \cite{necco-sommer} of the QCD potential 
between perturbative QCD predictions and a 
lattice computation.
More elaborated analyses on each of these comparisons
followed subsequently 
\cite{bsv2,Recksiegel:2001xq,Pineda:2002se,Lee:2002sn,impQCD}.
In all of these analyses, 
when IR contributions were appropriately eliminated,
the perturbative QCD predictions turned out to agree with the 
experimental data/phenomenological potentials/lattice results
within estimated perturbative uncertainties. 
Contrary to wide beliefs, there were no indications of 
large non--perturbative effects.
Only much smaller non--perturbative contributions, 
which can be absorbed into perturbative uncertainties,
appear to be compatible with these analyses.

In this paper we are concerned with the third type of comparison:
perturbative QCD prediction vs.\ lattice calculations.
In the previous comparisons of this type, 
the leading renormalon uncertainty of
the perturbative QCD potential was removed in various manners.
In \cite{necco-sommer}, 
the inter--quark force (improved by renormalisation group)
was used instead of the QCD potential\footnote{ 
See also \cite{Sumino:2001eh}
for some theoretical discussion on the inter--quark force.
};
in \cite{Pineda:2002se}, the leading renormalon contribution (estimated by a 
sophisticated approximation) was subtracted from the QCD potential
by hand;
in \cite{Lee:2002sn}, the perturbative series was Borel--resummed,
taking into account the leading renormalon pole appropriately.
We examine yet another method for removing the leading renormalon.
Namely, we examine the total energy $E_{\rm tot}(r)$, as defined above,
after re--expressing it in terms of
the $\overline{\rm MS}$ mass renormalised at the
$\overline{\rm MS}$ mass scale,
$\overline{m} \equiv m^{\overline{\rm MS}}(m^{\overline{\rm MS}})$.
To achieve stable predictions over a wide range of $r$, 
we adopt the scale--fixing prescriptions of 
\cite{Sumino:2001eh,Recksiegel:2001xq}.
These prescriptions introduce the scales
dependent on $r$ and \m, $\mu=\mu (r,\overline{m})$, which are
consistent with physical expectations.

As stated, $E_{\rm tot}(r)$ constitutes the leading part of 
the non--relativistic Hamiltonian of the heavy quarkonium system
within the potential--NRQCD framework.
We may expect that a direct comparison
of $E_{\rm tot}(r)$, which determines the bulk of the heavy
quarkonium spectrum,
would provide a clearer picture of the present status on
the credibility of the theoretical predictions based on
the potential--NRQCD framework,
supplemented either by perturbative QCD computations or by 
lattice computations of the potentials; see e.g.\
\cite{Brambilla,Bali,Poincare,impQCD} for analyses in this direction.
Furthermore, the scale--fixing prescription for $E_{\rm tot}(r)$ we adopt here 
is the only prescription which has been used in the 
perturbative QCD predictions for the level structure of the bottomonium states
(including higher excited states)
incorporating the renormalon cancellation.
We will see that indeed this prescription stabilizes the
perturbative prediction up to large distances, and hence it is suited
for predicting the energy levels of
excited states of the heavy quarkonium systems.

According to the renormalon argument, an uncertainty of the perturbative
QCD prediction for $E_{\rm tot}(r)$ grows rapidly at large distances
as $\Lambda_{\rm QCD}^3\, r^2$ \cite{al}.
It is nevertheless
important to predict $E_{\rm tot}(r)$ perturbatively at large $r$
for the following reasons.
(1)~The level spacings among the bottomonium
spectrum have uncertainties smaller than the uncertainties of 
individual levels.  
This is because the errors of $E_{\rm tot}(r)$ at 
different $r$ are generally correlated.  
Indeed, the estimate of the
error of $E_{\rm tot}(r)$, by changing input parameters or 
scale--fixing prescriptions, is perfectly consistent with 
$\Lambda_{\rm QCD}^3\, r^2$ \cite{Sumino:2001eh,Recksiegel:2001xq}; 
on the other hand, the bottomonium level spacings vary less, 
because the 
individual levels vary in a correlated way.  
This is why
the perturbative QCD predictions of the whole level structure of the
bottomonium in \cite{bsv1,bsv2,impQCD} made sense.
(2)~Many physical quantities of heavy quarkonium states are
sensitive to short--distance part of the potential.
For instance, the fine splittings of the bottomonium excited states 
are sensitive to much 
shorter distance part of $E_{\rm tot}(r)$ as compared to the individual 
levels. 
As a result, perturbative uncertainties of the fine splittings 
are much more suppressed (of order $\Lambda_{\rm QCD}^3/m^2$) as compared to 
uncertainties of the individual levels which directly reflect uncertainties 
of $E_{\rm tot}(r)$.  
Predictability of $E_{\rm tot}(r)$ up to large distances ensures 
that the wave functions of the excited states can be computed
in the computation of the fine splittings \cite{impQCD}, although only 
the short--distance parts of the wave functions are relevant.  
The order $\Lambda_{\rm QCD}^3\, r^2$
uncertainty of $E_{\rm tot}(r)$ at large distances is just
appropriate to ensure the theoretical uncertainties 
($\Lambda_{\rm QCD}^3/m^2$) 
of the fine splittings.  
These theoretical uncertainties, as well as the level of 
agreement with the experimental values, of the computed fine splittings (and 
the hyperfine splittings) turn out to be 
comparable to those of the recent lattice 
computations of these splittings; see \cite{impQCD,hyperfine} for details.  

In our comparison of the QCD potential
between perturbative QCD and lattice computations, 
we benefit from considering a hypothetical world which
contains no light quark flavours.
It is then possible to use the lattice calculations of the QCD potential
in the quenched approximation.
On the other hand, in the perturbative prediction for $E_{\rm tot}(r)$,
we have an additional parameter.
Although naively the quark mass is simply a constant independent of
$r$, due to our specific scale--fixing prescriptions,
the value of $\overline{m}$ affects the $r$--dependence of 
$E_{\rm tot}(r)$ non--trivially.
For a heavy quarkonium system in this hypothetical world, 
there is no strong motivation to choose a specific value for $\overline{m}$
(as opposed to the studies \cite{Sumino:2001eh,Recksiegel:2001xq}).
Therefore, in our analysis, we treat $\overline{m}$ as a controllable
parameter for testing stability of the perturbative prediction.
We will show that for those choices of \m\ that give stable predictions, 
$E_{\rm tot}(r)$ is independent of \m\ up to deviations of the order of the 
expected theoretical uncertainty (after a suitable shift by an 
$r$--independent constant). 
By varying \m\ to achieve optimum convergence for
large $r$, we can obtain perturbative QCD predictions 
up to fairly long distances and compare them
to the results of lattice QCD.

The organisation of the paper is as follows:
Sec.~\ref{perturbative} sets our conventions and gives some details of our
perturbative QCD calculation.
Sec.~\ref{comparison}
compares the lattice and perturbative QCD data.
Conclusions are given in Sec.~\ref{conclusions}.
We collect formulae related to the renormalisation--group evolution
of the strong coupling constant in the Appendix.

%%%%%%%%%%%%%%%%%%%%%%%%%%%%%%%%%%%%%%%%%%%%%%%%%%%%
\section{Conventions and framework} \label{perturbative}

We would like to 
compare the lattice data and the perturbative predictions 
corresponding to the same theoretical input.
This will be carried out in the following manner.
For each lattice data set we calculate the Sommer scale $r_0$ defined by
\bea  
r^2 \,{dV_{\rm QCD}\over dr} \bigg|_{r=r_0} = 1.65 . 
\label{def-Sommer}
\eea
Then the lattice data are expressed in units of $r_0$.
The perturbative computations are expressed 
in terms of the strong coupling constant defined in the
theory with $n_l=0$ active flavours.
We convert all the results into units of $r_0$
using the relation between the Lambda parameter of the 
running coupling constant (in the $\overline{\rm MS}$ scheme)
and the Sommer scale \cite{Capitani:1998mq}:
$\Lambda_{\overline{\rm MS}}=0.602(48)\,r_0^{-1}$. 
We use the central value of this relation in the main part of
our analysis; the effect of a variation of $\Lambda_{\overline{\rm MS}}$
inside the error interval is discussed at the end of Sec.~\ref{comparison}.
All the predictions are compared in units of $r_0$.
Furthermore, in order to maintain physical intuition, we 
will also use physical units.
Although there exists no rigid correspondence between the physical scales
of the real world and of the hypothetical world,
we follow the convention of the lattice
calculations in the quenched approximation.
The numerical value on the right--hand--side of Eq.~(\ref{def-Sommer})
has been chosen so that for phenomenological potentials 
$r_0 \approx (400\, {\rm MeV})^{-1}$.
Whenever we refer to values in units of MeV or GeV, we invoke
this translation.
\medbreak

The total energy of a static quark antiquark system is given by
\beq E_{\rm tot}(r)=2m_{\rm pole}+V_{\rm QCD}(r) \, . \label{Etot2} \eeq
In perturbative QCD, 
the pole mass \mpole\ is related to the $\overline{\rm MS}$
mass \m\ up to three loops by the relation
\beq
m_{\rm pole} = \overline{m} \left\{ 1 + {4\over 3}\,  
{\alpha_S(\overline{m})\over \pi} 
+ \left({\alpha_S(\overline{m})\over \pi}\right)^2 d_1 
+ \left({\alpha_S(\overline{m})\over \pi}\right)^3 d_2 \right\}
\label{massrel}.
\eeq
The QCD potential up to ${\cal O}(\alpha_S^3)$ is given by
\bea
V_{\rm QCD}(r) & = &
- \, \frac{4}{3} \frac{\alpha_S(\mu)}{r} \Biggl[ \, 1 + 
\biggl( \frac{\alpha_S(\mu)}{4\pi} \biggr) \,
  ( 2 \beta_0 \ell + a_1 ) \nonumber \\
&&\hspace{2.4cm}+ \biggl( \frac{\alpha_S(\mu)}{4\pi} \biggr)^2
  \left\{ \beta_0^2 
  \Bigl( 4 \ell^2  + \frac{\pi^2}{3} \Bigr)
  + 2 ( \beta_1 + 2 \beta_0 a_1 ) \ell + a_2 \right\} \Biggr] ,
\label{QCDpot}
\eea
where $\ell = \log (\mu r) + \gamma_E $.

Here and hereafter, we have
set the number of light flavours, $n_l$, to zero, i.e.\
we will be neglecting the effects of light quark loops. This
corresponds to the quenched approximation in lattice QCD to
which we want to compare the perturbative results. The running
coupling $\alpha_S(\mu)$ depends on $n_l$ through the coefficients
of the beta function; the constants $d_1, d_2, a_1$ and $a_2$
also get contributions from light quark loops and therefore
depend on $n_l$. 
For $n_l=0$, their values are $\beta_0=11,\,
\beta_1=102,\, d_1\approx 13.443,\, d_2\approx 190.39,\, a_1=31/3$ 
and $a_2\approx 456.74$. The analytical formulae can be found in
\cite{Recksiegel:2001xq}.\footnote{
These formulae have originally been computed in \cite{ps,mr}.
The mass relation (\ref{massrel}) is
re--expressed in terms of the coupling of the theory without
heavy quarks.
}

After re--expressing $\alpha_S(\overline{m})$ in Eq.~(\ref{massrel}) in
terms of $\alpha_S(\mu)$ by using the running of $\alpha_S$
[see Eq.~(\ref{pertrel}) in Appendix], and
dropping terms of ${\cal O}(\alpha_S(\mu)^4)$ and higher, we obtain
the total energy $E_{\rm tot}(r;\overline{m},\alpha_S(\mu),\mu)$ 
which does not
suffer from the leading renormalon uncertainty.

Due to the truncation of the perturbative series at finite order,
$E_{\rm tot}$ depends on the renormalisation scale
$\mu$. Two scale fixing prescriptions 
have been introduced in \cite{Sumino:2001eh}:
\begin{enumerate}
\item
The scale $\mu = \mu_1(r)$ is fixed by demanding stability 
of $E_{\rm tot}(r)$ against variation of the scale:
\bea \mu \frac{d}{d\mu} E_{\rm tot}(r;\overline{m},\alpha_S(\mu),\mu)
\,\bigg|_{\mu = \mu_1(r)} = 0 .
\label{scalefix1} \eea
\item
The scale $\mu = \mu_2(r)$ is fixed to the minimum of the absolute 
value of the last known term [${\cal O}(\alpha_S^3)$ term] of $E_{\rm tot}(r)$:
\bea \mu \frac{d}{d\mu}
\Bigl[ E_{\rm tot}^{(3)}(r;\overline{m},\alpha_S(\mu),\mu)\Bigr]^2 \,
\,\bigg|_{\mu = \mu_2(r)} = 0 .
\label{scalefix2}
\eea
\end{enumerate}
Although these prescriptions are very different, it has been 
shown that where both prescriptions exist, the total energy is 
virtually identical for both prescriptions.
As a general feature of $E_{\rm tot}(r)$,
the convergence of the perturbative series improves and the
scale dependence decreases, if we choose larger $\mu$
for smaller distances and smaller $\mu$ for larger distances.
Consequently,
the range of the perturbative calculation can be extended 
to much larger $r$ with these prescriptions than what 
would be possible with a fixed, $r$--independent scale.

The prescriptions for the renormalon cancellation and the scale
fixing we adopt here follow (basically)
those in \cite{Sumino:2001eh,Recksiegel:2001xq,impQCD}.
We refer the reader to these papers for more detailed features
of the perturbative predictions in these prescriptions.

There are several methods to assess the reliability of the
prediction at a given distance: e.g.\ one can compare
the total energies as determined with the scales from both 
prescriptions, one can
compare the sizes of the individual terms of the perturbative
expansion of \Etot\ or one can study the scale dependence of
\Etot\ around the respective scale. We will use these methods
in section \ref{comparison}.

\begin{table}
\begin{center}
\begin{tabular}{|c||ccrr|c||ccrc|c|}
\hline
  \multirow{2}{.3cm}{$\overline{m}$} 
 & \multicolumn{5}{|c||}{$\mu=\mu_1$} & \multicolumn{5}{|c|}{$\mu=\mu_2$} \\
 & $\mu$ & $E_{\rm tot}^{(1)}$ & $E_{\rm tot}^{(2)}$  & $E_{\rm tot}^{(3)}$ & $E_{\rm tot}$ 
 & $\mu$ & $E_{\rm tot}^{(1)}$ & $E_{\rm tot}^{(2)}$  & $E_{\rm tot}^{(3)}$ & $E_{\rm tot}$ \\
\hline
 $1.6$ & $0.389$ & $1.275$ & $0.271$ & $-0.280$ & $4.466$ & $0.419$ & $0.921$ &  $0.243$ & $0$ & $4.364$ \\
 $1.8$ & $0.413$ & $1.126$ & $0.109$ & $-0.147$ &  $4.687$ & $0.449$ & $0.881$ & $0.158$ & $0$ & $4.639$ \\
 $2.0$ & $0.436$ &$1.08$ &  $0.038$ & $-0.096$ & $5.022$ & $0.477$ & $0.882$ & $0.111$ & $0$ &  $4.993$ \\
 $2.2$ & $0.458$ & $1.073$ & $-0.007$ & $-0.069$ & $5.397$ &$0.502$ &  $0.901$ & $0.077$ & $0$ & $5.378$ \\
 $2.4$ & $0.478$ & $1.085$ & $-0.042$ &  $-0.051$ & $5.792$ & $0.525$ & $0.929$ & $0.049$ & $0$ & $5.778$ \\
 $2.6$ & $0.497$ & $1.109$ & $-0.072$ & $-0.039$ & $6.197$ & $0.545$ & $0.965$ &$0.022$ &  $0$ & $6.187$ \\
 $2.8$ & $0.515$ & $1.140$ & $-0.102$ & $-0.029$ & $6.609$ &  $0.563$ & $1.006$ & $-0.005$ & $0$ & $6.601$ \\
 $3.0$ & $0.530$ & $1.179$ &  $-0.133$ & $-0.021$ & $7.025$ & $0.576$ & $1.055$ & $-0.035$ & $0$ &  $7.019$ \\
 $3.2$ & $0.543$ & $1.224$ & $-0.168$ & $-0.012$ & $7.444$ &$0.453$ &  $1.680$ & $-0.682$ & $0$ & $7.398$ \\
 $3.4$ & $0.553$ & $1.277$ & $-0.208$ &  $-0.004$ & $7.865$ & $0.507$ & $1.451$ & $-0.393$ & $0$ & $7.859$ \\
 $3.6$ & $0.559$ & $1.342$ & $-0.262$ & $0.006$ & $8.287$ & $0.577$ & $1.288$ &$-0.207$ &  $0.005$ & $8.287$ \\
 $3.8$ & $0.554$ & $1.439$ & $-0.352$ & $0.023$ & $8.711$ &  $0.615$ & $1.268$ & $-0.173$ & $0.012$ & $8.708$\\
\hline
\end{tabular}
\caption{Convergence properties of $E_{\rm tot}$ for $r=2r_0\simeq
  5\,{\rm GeV}^{-1}$. All numbers in GeV. \label{convergence}}
\end{center}
\end{table}
The convergence properties of $E_{\rm tot}(r)$ strongly depend on the
mass parameter \m. 
We illustrate this in Table \ref{convergence}
for $r=2r_0\simeq 5\,{\rm GeV}^{-1}$.
$E_{\rm tot}^{(i)}$ denotes the ${\cal O}(\alpha_S^i)$ term of the
perturbative series of $E_{\rm tot}$. It can be seen that the
series converges nicely, especially for $\overline m \simeq 3\,
{\rm GeV}$. In section \ref{comparison} we will see that those
values for \m\ that provide best convergence also provide an
optimal agreement with lattice results.
We find that, to our surprise, the perturbative series converges for
distances as large as $3r_0\simeq7.5\,{\rm GeV}^{-1}$. 
(Generally, the convergence behaviour of $E_{\rm tot}(r)$
becomes worse for larger $r$.)
We note that the values of the scales $\mu_{1,2}$ stay much larger
than $1/r$.
See \cite{Sumino:2001eh} for discussion on this aspect.

\begin{figure}
\begin{center}
\psfrag{r}{$r/r_0$} \psfrag{V}{$r_0\cdot V+{\rm const}.$}
\psfrag{analytical running}{analytical running}
\psfrag{numerical running}{numerical running}
\includegraphics[width=15cm]{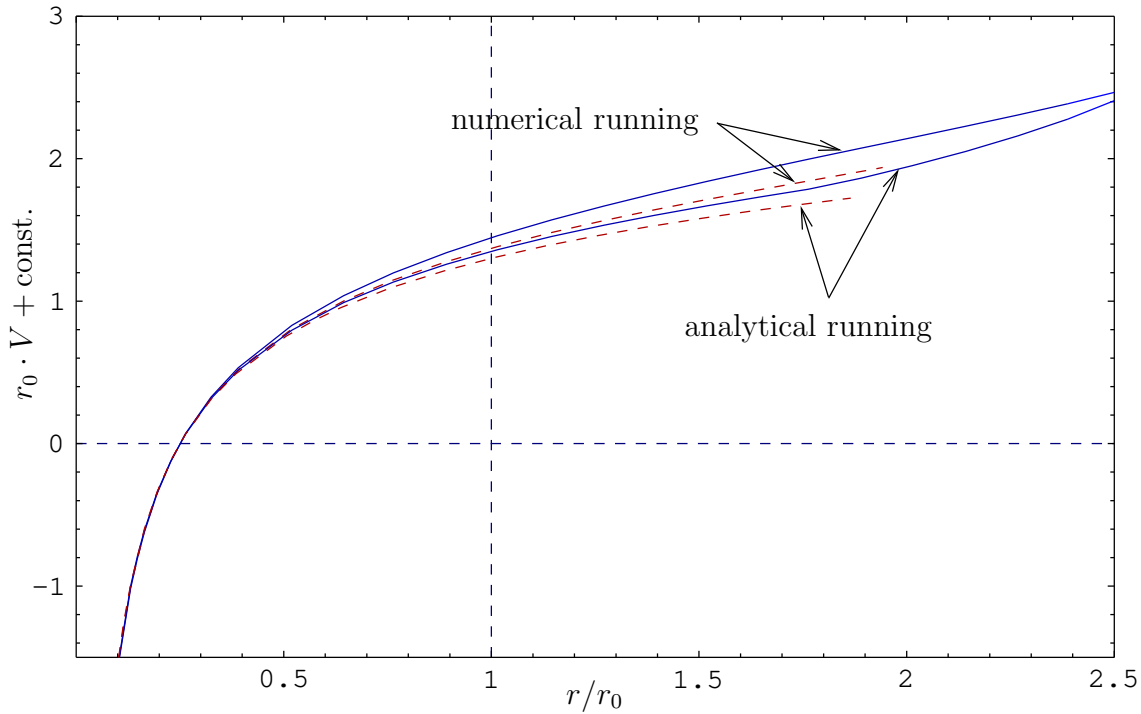}
\end{center}
\vspace*{-.5cm}
\caption{Comparison between analytical and numerical running. The
solid lines correspond to $\overline m =3 \,{\rm GeV}\simeq 7.5 /r_0$, the
dashed ones to $\overline m =4 \,{\rm GeV}\simeq 10 /r_0$.
\label{ananumcomparison}}
\end{figure}
For large distances, $E_{\rm tot}$ depends on whether we use the
analytical solution or numerical solution of the renormalisation--group
equation for the running of the strong coupling constant 
$\alpha_S(\mu)$ (see Appendix).
We show the total energy for both types of running in 
Fig.~\ref{ananumcomparison} for two values of \m. In accordance with
our previous works \cite{Sumino:2001eh,Recksiegel:2001xq} we
employ the numerical solution of the renormalisation--group
equation below. 
Our results do not change qualitatively if we
use the analytical running instead.
\medbreak

To end this section, 
we briefly summarise the argument on the renormalons included in
$E_{\rm tot}(r)$.
\begin{itemize}
\item[(1)]
After the ${\cal O}(\Lambda_{\rm QCD})$ 
renormalon is cancelled, the perturbative series 
of $E_{\rm tot}(r)$ is estimated to have the following behaviour.  
The ${\cal O}(\alpha_S^{n+1})$ term of the series 
expansion of $E_{\rm tot}(r)$ behaves as 
${ const}.\times r^2 \, n! \, (\beta_0 \alpha_S/(6\pi))^n \, n^{3\delta/2}$
for $n \gg 1$, where $\delta = \beta_1/\beta_0^2$.
Because of the factorial $n!$, the series is only 
an asymptotic series, namely it diverges for large enough $n$.  
Hence, 
there is a limitation to the achievable accuracy of the perturbative 
prediction for $E_{\rm tot}(r)$.  
It can be estimated from the size of the terms 
around the minimum, 
$n \approx 6\pi/(\beta_0 \alpha_S)$;
this gives an uncertainty of order 
$\Lambda_{\rm QCD}^3r^2$.  
The behaviour of the series depends on the value of the 
expanding parameter $\alpha_S \equiv \alpha_S(\mu)$, or equivalently, 
on the choice of the scale $\mu$.  
The uncertainty, $\Lambda_{\rm QCD}^3r^2$, is independent of 
$\alpha_S(\mu)$ or $\mu$, nonetheless.  
(For details, see \cite{beneke,Sumino-pro}.)

\item[(2)] 
Based on the above argument, we may optimise convergence of the 
series by appropriately choosing the scale $\mu$.
In this case, even with the series expansion
up to ${\cal O}(\alpha_S^3)$ we may estimate the uncertainty of 
$E_{\rm tot}(r)$ to be of order $\Lambda_{\rm QCD}^3r^2$ from the
size of
truncated next-order term or from the scale dependence of
$E_{\rm tot}(r)$ around the optimised scale.  
Indeed, explicit numerical examinations of $E_{\rm tot}(r)$ 
up to ${\cal O}(\alpha_S^3)$ support this argument 
\cite{Sumino:2001eh,necco-sommer,Recksiegel:2001xq,Pineda:2002se}.

\item[(3)] 
One may factorise the infrared part of $E_{\rm tot}(r)$ using 
operator product expansion \cite{bpsv}.  
In this way, one may absorb the order $\Lambda_{\rm QCD}^3r^2$
renormalon into a matrix 
element of an operator, while defining a Wilson coefficient perturbatively 
that is free from the renormalon and dependent on the factorisation 
scale.  
More generally, one may separate $E_{\rm tot}(r)$ into perturbative 
coefficients free from renormalons and non--perturbative parameters
(matrix elements) including renormalons.  
(This factorisation is beyond the scope of the present paper.)

\end{itemize}

%%%%%%%%%%%%%%%%%%%%%%%%%%%%%%%%%%%%%%%%%%%%%%%%%%%%
\section{Comparison of perturbative and lattice calculations} 
\label{comparison}

For comparison with the perturbative predictions of the QCD potential
as explained in the previous section,
four different sets of lattice data calculated in the quenched approximation
are used:
those from \cite{Bali:1992ru} ($\beta=6.8$), from \cite{suganuma} ($\beta=6.0$),
from \cite{kaneko} ($\beta=6.0$) and from 
\cite{necco-sommer-data} ($6.57\le \beta\le 6.92$).
All the lattice data
have been corrected using the lattice Coulomb potential 
to match the continuum definition of
the QCD potential at short distances.

%All our plots are normalised to $r_0$ and therefore scaleless,
%but $r_0^{-1}=400{\rm MeV}$ has been used where dimensionful
%quantities are quoted (e.g.\ masses).
For comparison of the perturbative and the lattice 
data, we have to account for an $r$--independent 
additive constant that is not determined by
lattice calculations. Since perturbative calculations are more
reliable at small distances and lattice calculations are more
reliable at larger distances, we adopt the following procedure:
The different sets of lattice data are converted to physical
units with the lattice spacing as given by the authors of the
respective papers, or (where the lattice spacing was not explicitly
derived) by fixing the Sommer scale with the phenomenological potential
fit as performed by the authors of the respective papers. Then we
adjust the sets of lattice data among each other to make them
coincide at $r=r_0$ by adding constants. 
Finally we shift both the perturbative and
the lattice data so that they vanish at $r=r_0/4$, where in the
case of the lattice results the data from \cite{necco-sommer-data} is 
used.

\begin{figure}
\begin{center} 
\psfrag{B}{$\circ$} \psfrag{K}{$\star$} \psfrag{S}{$\diamond$}
\psfrag{N}{$\bullet$} \psfrag{C}{$\bullet$}
\psfrag{Bali}{Bali, Schilling}\psfrag{Kaneko}{JLQCD}
\psfrag{Suganuma}{Takahashi et al.}\psfrag{Necco}{Necco, Sommer}
\psfrag{m=1}{$\overline m=1\,{\rm GeV}$}\psfrag{m=2}{$\overline m=2\,{\rm GeV}$}
\psfrag{m=3}{$\overline m=3\,{\rm GeV}$}\psfrag{m=4}{$\overline m=4\,{\rm GeV}$}
\psfrag{m=5}{$\overline m=5\,{\rm GeV}$}\psfrag{m=6}{$\overline m=6\,{\rm GeV}$}
\psfrag{r}{$r/r_0$} \psfrag{V}{$r_0\cdot V+{\rm const}.$}
\includegraphics[width=16cm]{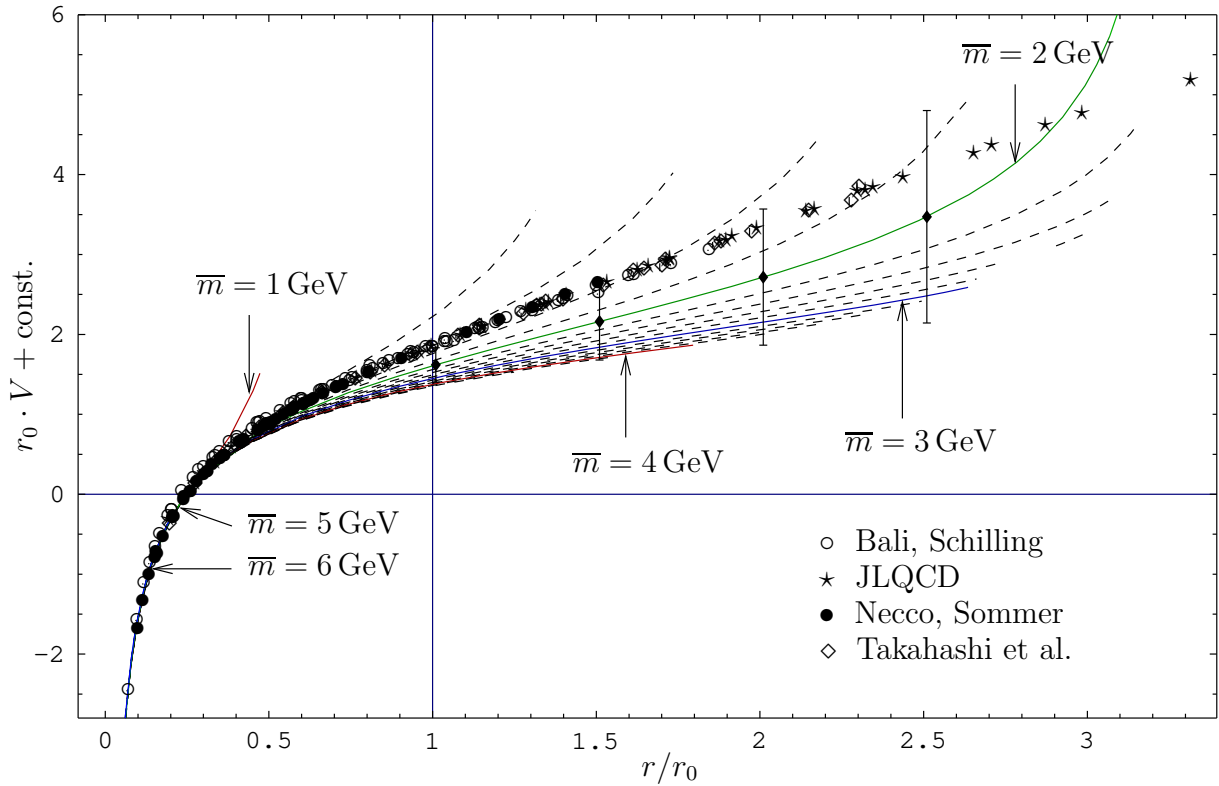}
\end{center}
\vspace*{-.5cm}
\caption{Comparison between perturbative and lattice calculations
of the QCD potential. The lines correspond to masses between 1
and 6 GeV in steps of 0.2 GeV
(solid lines for integer masses, $1 \,{\rm GeV}\simeq 2.5 /r_0$,
 the lines for $\overline m \simgt 4\,{\rm GeV}$ are masked by the other lines).
The points correspond to Bali/Schilling ($\circ$) \cite{Bali:1992ru}, 
Takahashi et al.\ ($\diamond$) \cite{suganuma}, JLQCD ($\star$) \cite{kaneko} 
and Necco/Sommer ($\bullet$) \cite{necco-sommer-data}.
Error bars for the statistical errors of the lattice data are 
plotted where given by the authors, but they are generally
smaller than the size of the symbols. Lines are plotted only when the
total energies determined by the two prescriptions differ by
less than $0.5/r_0$.
\label{mu1mu2num}}
\end{figure}

We see that the sets of lattice data 
\cite{Bali:1992ru,suganuma,kaneko,necco-sommer-data}
corresponding to different 
values of $\beta$ are located almost on the
same curve (Fig.~\ref{mu1mu2num}).
This shows that the dependence of
the lattice results on the lattice spacing is negligibly small,
i.e.\ discretization errors in the lattice calculations 
are negligible in our comparison.

As described before, the perturbative calculation has two input
parameters, these can be e.g.\ $\Lambda_{\overline{\rm MS}}$ and \m.
The potential depends on the mass \m\ (after shifting to
$E_{\rm tot}(r_0/4)=0$) only through the 
$\log ({\overline m}/\mu$)--terms in the relation between the pole
mass and the $\overline{\rm MS}$ mass. We find
that (after shifting the curves to make them coincide at $r_0/4$),
for small distances $r<r_0/4$ the curves are identical, for larger
distances the curves corresponding to different \m\ start
to differ from each other. The quality of convergence and the
stability against scale changes varies strongly with \m. 

In Fig.~\ref{mu1mu2num} we plot the QCD potential for various
values of \m\ between 1 and 6 GeV in steps of 0.2 GeV. 
To ensure
that only reasonably stable and reliable predictions are shown, the curves
are drawn only in those points, where the energies as determined
by the two different scale fixing prescriptions differ by less
than $0.5/r_0$. We find that
the resulting curves span a band around the lattice data
that increases in width with increasing $r$.
The width of this band
is consistent with the expected theoretical uncertainty due
to the uncancelled next--to--leading renormalon \cite{al}, 
$\pm \frac{1}{2} \Lambda^3 r^2$, with $\Lambda=300\,{\rm MeV}$, indicated
by the error bars in the figure.
We find a very good agreement between the lattice results
and the curves that show the largest range of convergence,
but even for those choices of \m\ where the prediction
becomes unstable earlier, the agreement is still good.

\begin{figure}
\begin{center}
\psfrag{B}{$\circ$} \psfrag{K}{$\star$} \psfrag{S}{$\diamond$}
\psfrag{N}{$\bullet$} \psfrag{C}{$\bullet$}
\psfrag{Bali}{Bali, Schilling}\psfrag{Kaneko}{JLQCD}
\psfrag{Suganuma}{Takahashi et al.}\psfrag{Necco}{Necco, Sommer}
\psfrag{m=1}{$\overline m=1\,{\rm GeV}$}\psfrag{m=2}{$\overline m=2\,{\rm GeV}$}
\psfrag{m=3}{$\overline m=3\,{\rm GeV}$}\psfrag{m=4}{$\overline m=4\,{\rm GeV}$}
\psfrag{m=5}{$\overline m=5\,{\rm GeV}$}\psfrag{m=6}{$\overline m=6\,{\rm GeV}$}
\psfrag{r}{$r/r_0$} \psfrag{V}{$r_0\cdot V+{\rm const}.$}
\includegraphics[width=16cm]{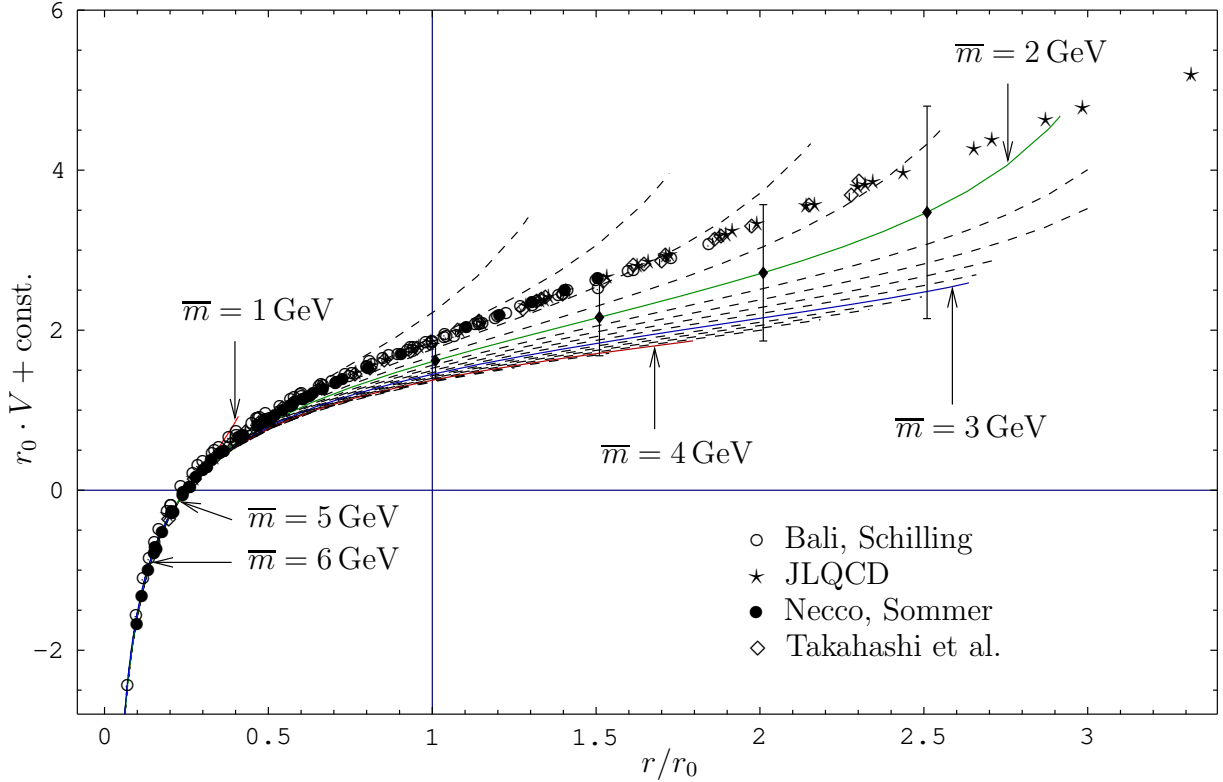}
\end{center}
\vspace*{-.5cm}
\caption{Comparison between perturbative and lattice calculations
of the QCD potential. The lines correspond to masses between 1
and 6 GeV in steps of 0.2 GeV. Lines are plotted only
when a +10\% scale change makes the total energy vary by less
than 20\%.
\label{varyscalenum}}
\end{figure}
To show that the good agreement between the perturbative and
the lattice calculations does not depend on a specific stability
criterion, in Fig.~\ref{varyscalenum} we show the same comparison as
in Fig.~\ref{mu1mu2num}, but this time we do not consider the
difference between the energies as determined with the two different 
scale prescriptions, but the stability against scale change.
We plot the curves only in those points where a scale change of +10\%
makes the total energy vary by less than $\pm$20\%.

We would like to stress that we do not tune the mass parameter
to achieve good agreement with the lattice results, but we
vary it to find those values of \m\ that give optimal convergence
of the perturbative series. It can be seen in the figures that
the curves for those values of \m\ that have the largest range
of convergence, the agreement with the lattice data is close
to optimal.

\begin{figure}
\begin{center}
\psfrag{m=1}{$\overline m=1\,{\rm GeV}$}\psfrag{m=4}{$\overline m=4\,{\rm GeV}$}
\psfrag{m=7}{$\overline m=7\,{\rm GeV}$}\psfrag{m=10}{$\overline m=10\,{\rm GeV}$}
\psfrag{Pineda}{$\mu={\rm const}.$}
\psfrag{r}{$r/r_0$} \psfrag{V}{$r_0\cdot V+{\rm const}.$}
\includegraphics[width=15cm]{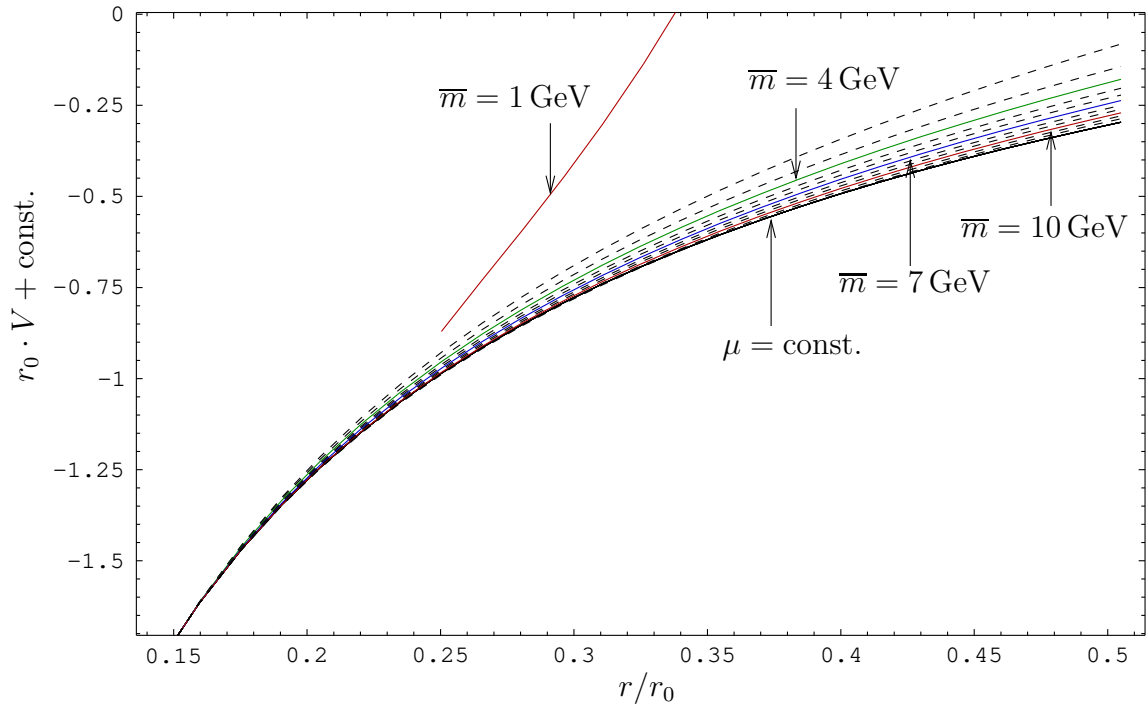}
\end{center}
\vspace*{-.5cm}
\caption{Comparison between our formalism and the 2--loop
QCD potential according to the formalism of \cite{Pineda:2002se}.
The lowermost line shows the (\m\ independent) potential for
a constant $\mu=(0.15399\, r_0)^{-1}$, corresponding to the formalism 
of \cite{Pineda:2002se}. The other lines show our results for masses 
from 1 to 12 GeV in steps of 1 GeV (solid lines for 1, 4, 7 and 10 GeV).
\label{comparetopineda}}
\end{figure}

\begin{figure}
\begin{center}
\psfrag{m=1}{$\overline m=1\,{\rm GeV}$}\psfrag{m=4}{$\overline m=4\,{\rm GeV}$}
\psfrag{m=7}{$\overline m=7\,{\rm GeV}$}\psfrag{m=10}{$\overline m=10\,{\rm GeV}$}
\psfrag{r}{$r/r_0$} \psfrag{mu}{$\mu_1(r)$}
\includegraphics[width=15cm]{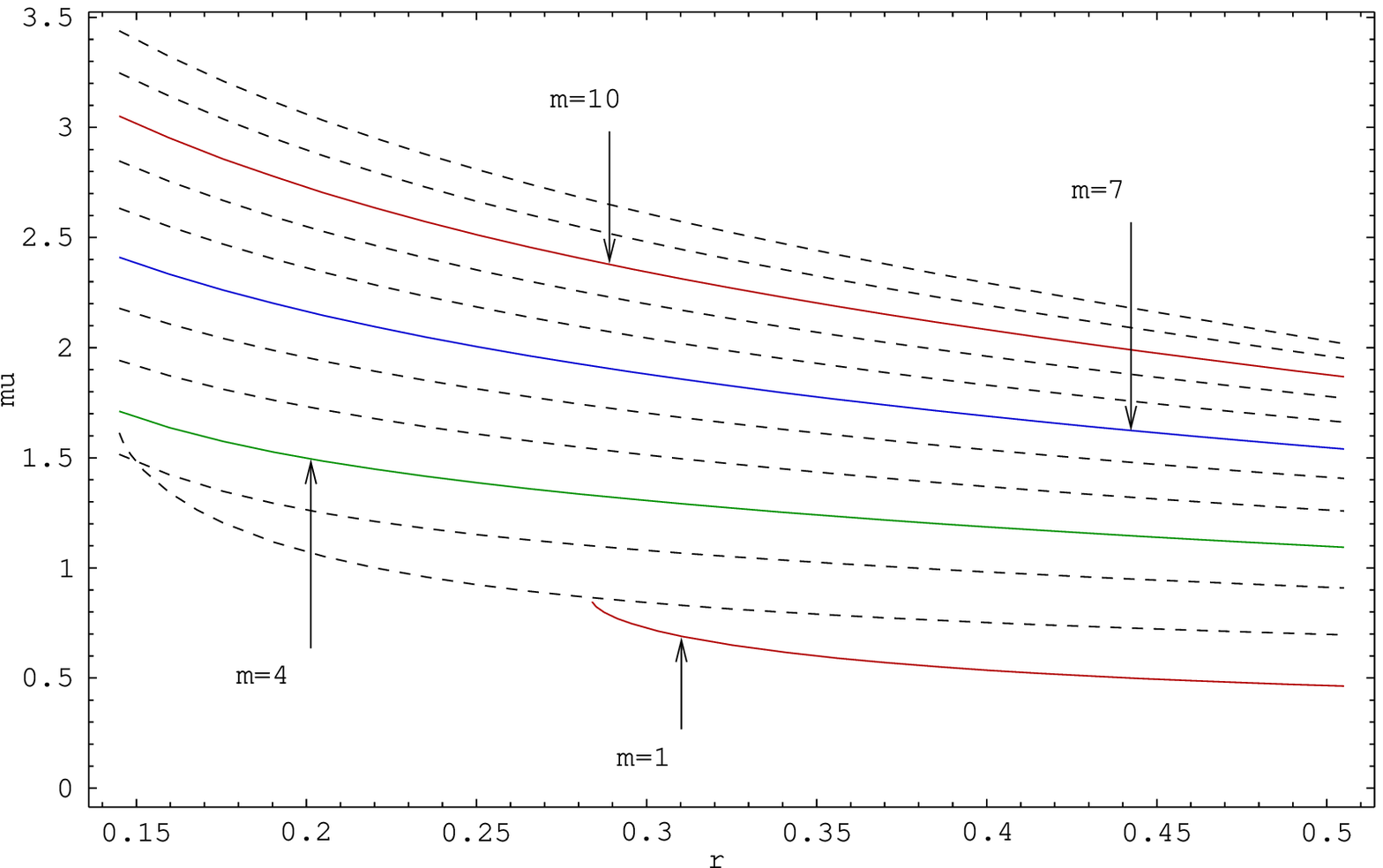}
\end{center}
\vspace*{-.5cm}
\caption{Scales as determined by the first prescription
(\ref{scalefix1}) for values of \m\ between 1 and 12 GeV. \label{scales}}
\end{figure}

We now compare our results to those of \cite{Pineda:2002se}.
In that paper, a fixed, $r$--independent scale $\mu$ is used for the
perturbative QCD potential. We find that our formalism almost
exactly reproduces the curves of \cite{Pineda:2002se} for large
values of \m\ (Fig.~\ref{comparetopineda}).
The explanation for this behaviour is the following: While in
our formalism the scale is strongly dependent on $r$ even for
large masses (see Fig.~\ref{scales}),
the scale tends to rise with \m. For $\overline{m} \sim 3 \,{\rm GeV}$
the $r$--dependent scale varies around 1 GeV, for $\overline m\sim 10 \,{\rm GeV}$
it varies around 3 GeV. Independent of \m, however, the scale
dependence of $E_{\rm tot}$ is strong for scales around 1 GeV and
\begin{figure}
\begin{center}
\psfrag{m=1}{$\overline m=1\,{\rm GeV}$}\psfrag{m=4}{$\overline m=4\,{\rm GeV}$}
\psfrag{m=7}{$\overline m=7\,{\rm GeV}$}\psfrag{m=10}{$\overline m=10\,{\rm GeV}$}
\psfrag{mu}{$\mu$}\psfrag{V}{$r_0\cdot V+{\rm const}.$}
\includegraphics[width=15cm]{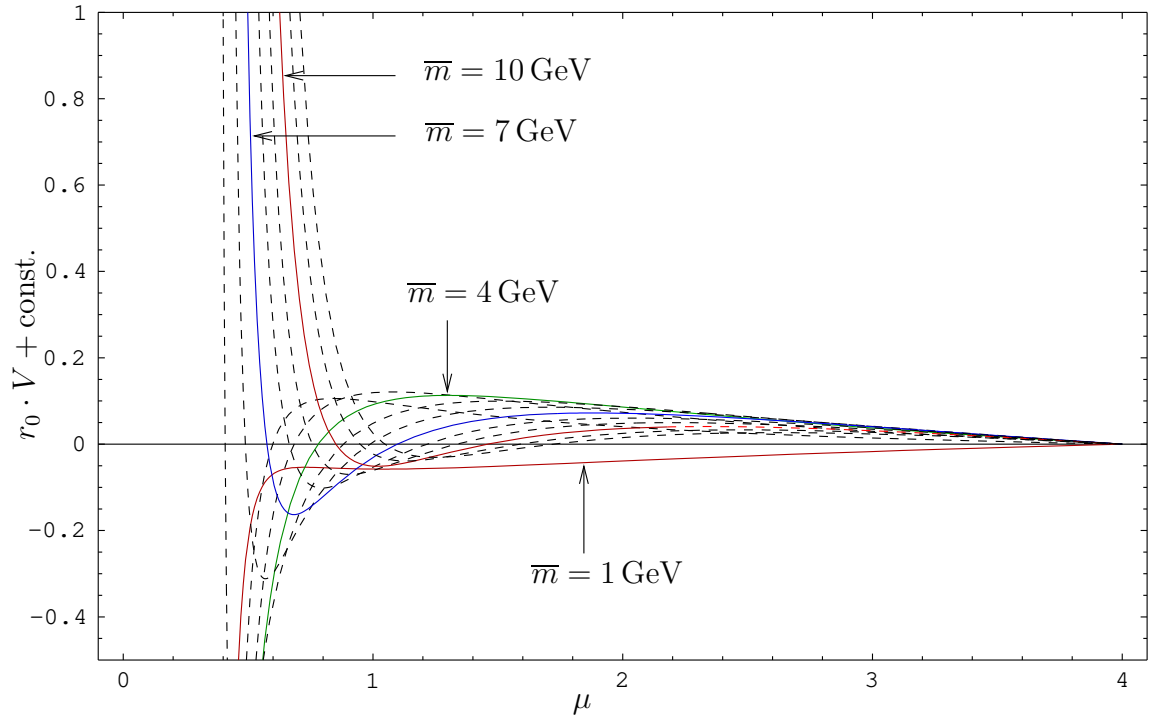}
\end{center}
\vspace*{-.5cm}
\caption{Dependence of $E_{\rm tot}(r=0.3\,r_0)$ on the scale $\mu$ for
values of \m\ between 1 and 15 GeV. Independent of \m, the scale
dependence is strong for $\mu\simlt 1\,{\rm GeV}$ and weak for
$\mu\simgt 1\,{\rm GeV}$.
\label{mudependence}}
\end{figure}
very weak for scales around 3 GeV (Fig.~\ref{mudependence}), 
therefore choosing a large \m\ in our formalism gives a result very 
close to the treatment of \cite{Pineda:2002se}. We can also see that
for these large masses our stability criteria indicate a range of
convergence up to about $r_0/2$. Our analysis is therefore 
consistent with that of \cite{Pineda:2002se}, and the results of the
latter are reproduced with our formalism by choosing large values
for \m.

\begin{figure}
\begin{center}
\psfrag{B}{$\circ$} \psfrag{K}{$\star$} \psfrag{S}{$\diamond$}
\psfrag{N}{$\bullet$} \psfrag{C}{$\bullet$}
\psfrag{Lambda}{$\Lambda_{\overline{\rm MS}}\,r_0=$}
\psfrag{L554}{$0.554$}\psfrag{L602}{$0.602$}\psfrag{L650}{$0.650$}
\psfrag{r}{$r/r_0$} \psfrag{V}{$r_0\cdot V+{\rm const}.$}
\includegraphics[width=15cm]{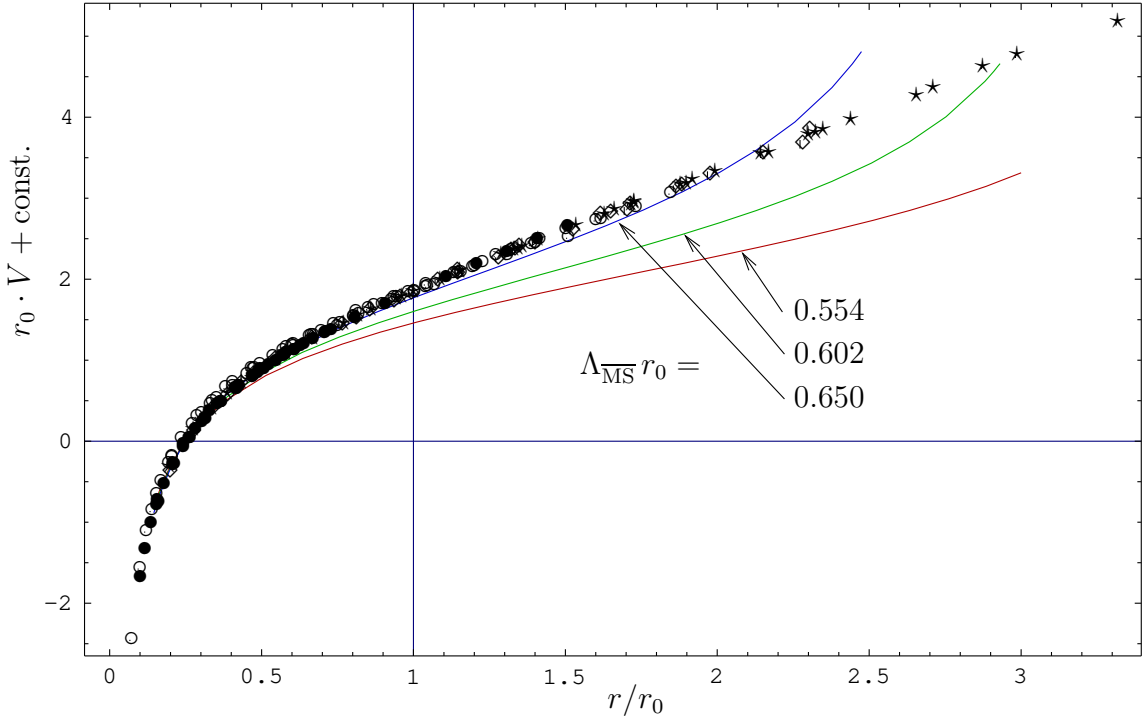}
\end{center}
\vspace*{-.5cm}
\caption{Dependence of the perturbative results on $\Lambda_{\overline{\rm
  MS}}r_0$, the curves shown correspond to the centre and upper and lower
  bounds of the error interval given in \cite{Capitani:1998mq}. The
  middle curve corresponds to the one in Fig.~\ref{varyscalenum} that
  has the error bars attached to it ($\overline m = 2\, {\rm GeV}$).
\label{varyLambdaMS}}
\end{figure}

The perturbative predictions which are most stable at long distances
in Figs.~\ref{mu1mu2num} and \ref{varyscalenum}
($\overline{m} \sim 2$--3~GeV) turn out to be less steep than the lattice data
at $r/r_0 \simgt 0.5$.
Qualitatively, we expect that the larger the strong coupling constant
the steeper the potential, because the interquark force
becomes stronger \cite{Sumino:2001eh}. 
%%%% completely new
This behaviour can be seen in Fig.~\ref{varyLambdaMS}
where we have varied $\Lambda_{\overline{\rm MS}}\,r_0$
in the interval given in \cite{Capitani:1998mq},
$\Lambda_{\overline{\rm MS}}=(0.602\pm 0.048)\,r_0^{-1}$. The lower bound,
centre and upper bound of this interval correspond to
$\alpha_S^{n_l=0}(M_Z)=0.0801,\, 0.0811$ and $0.08205$, respectively.
The larger value for $\Lambda_{\overline{\rm MS}}$ 
(if $r_0 \approx 2.5~{\rm GeV}^{-1}$ is fixed)
results in a
slightly steeper curve that reproduces the slope of the lattice
data better than the central value.
%%%%%%%%%%%%%%

We would like to make two comments in this context:
(i) We compared the perturbative QCD potential (including
effects of light quark loops) with phenomenological potentials in
\cite{Recksiegel:2001xq}.
There, the perturbative prediction with the input 
$\alpha_S(M_Z)=0.1181$ (the present central value) turned out
also to be slightly less steep than the phenomenological potentials 
at long distances.
As a result, a somewhat larger coupling $\alpha_S(M_Z)=0.1191$--$0.1201$
was favoured for a better agreement with the phenomenological potentials.
(ii) In \cite{Pineda:2002se}, the ${\cal O}(\alpha_S^4)$ correction
to the perturbative QCD potential (including the ultrasoft effects)
was estimated and included in a comparison with the lattice data.
The estimated correction makes the potential somewhat steeper.
This is consistent with a naive expectation that such an effect
is caused by an acceleration of the running of the coupling constant
due to the 4--loop coefficient of the beta function.
Thus, agreement of the perturbative potential 
with the lattice data may become even 
better than our present analysis when the full next order correction
is calculated and included in the future.
Furthermore, we have confirmed that the agreement between the
perturbative and lattice data becomes even better if we use the
4--loop running of the coupling constant.

\section{Conclusions} \label{conclusions}

We have compared perturbative QCD and lattice QCD predictions
for the QCD potential.
We examined the perturbative QCD prediction for
$E_{\rm tot}(r) = 2 m_{\rm pole} + V_{\rm QCD}(r)$,
taking specific prescriptions for fixing the
renormalisation scale $\mu$.
We find that, by adjusting the mass parameter \m, the perturbative 
prediction can be made stable up to distances $r \sim 3\, r_0$.
Whenever we obtain stable perturbative predictions, they agree with
the lattice data within the uncertainty estimated from
the residual renormalon of order $\Lambda^3 r^2$.
We emphasise that we do not tune \m\ to fit the lattice data,
but we tune \m\ to achieve stability of the perturbative prediction,
and then the agreement follows.
Comparisons of perturbative QCD predictions and lattice
data have been performed before e.g.\ in \cite{Pineda:2002se, Lee:2002sn},
but only up to distances of $0.5\, r_0$ and $0.9\, r_0$, respectively.
If we take an optimal value of \m, 
our prescriptions for the perturbative prediction of the QCD potential
seem to give stable predictions to furthest distances among those
examined so far.

Our analysis provides a firmer ground to the analyses of 
\cite{bsv1,bsv2,impQCD}, which predicted the bottomonium energy levels
up to the $n=3$ states using (essentially) the same scale--fixing prescriptions
as in the present analysis.\footnote{
The value of \m, which stabilizes
the perturbative predictions for $E_{\rm tot}(r)$ 
up to the furthest distance, lies between the bottom and charm quark masses.
In this sense, we are in a lucky situation in the predictions of
the charmonium and bottomonium spectra.
}
We note that the same conclusion could not be drawn directly from the previous
comparisons \cite{necco-sommer,Pineda:2002se,Lee:2002sn}
between the lattice and perturbative computations of the QCD potential,
because the prescriptions adopted in those analyses have never been
used in perturbative computations of the heavy quarkonium level structure
including higher excited states.
In the light of our present result, the scale--fixing prescription
adopted here is optimal for stable predictions for the energy levels of 
excited states. Our result supports the estimates of theoretical uncertainties
by the next--to--leading renormalon made in \cite{bsv1,bsv2,impQCD}.

Whenever stable perturbative 
predictions are obtained, all the perturbative predictions with
different prescriptions for subtracting the leading renormalon 
agree with one another and also with the lattice data, within
the estimated uncertainty.
In particular, our perturbative predictions for large \m\ reproduce the 
${\cal O}(\alpha_S^3)$ perturbative prediction of \cite{Pineda:2002se}.
The fact that the different prescriptions have 
led to mutually consistent perturbative
predictions of the QCD potential, endorses 
the consistency of the perturbative analyses.
The comparisons between the perturbative and lattice data, 
together with other types of comparisons, provide evidence to the
hypothesis that the perturbative prediction agrees with full QCD
within the order $\Lambda^3 r^2$ uncertainty.
In particular, we consider the independent
examinations corresponding to
the physical reality ($n_l=4$ with the non--zero
charm quark mass \cite{Recksiegel:2001xq}) 
and to the hypothetical case ($n_l=0$)
to be non--trivial cross checks with respect to the validity of the hypothesis.
We may further make non--trivial tests of the
hypothesis by varying the number of quark flavours and even the
gauge group in comparing the perturbative and lattice calculations.

It is quite surprising that the perturbative calculations turn
out to give stable predictions up to such long distances
[$3r_0 \approx (130~{\rm MeV})^{-1}$].
At the present stage the reason is unclear.
Although we do not know a clear--cut criterion at which
distance a perturbative 
QCD prediction should break down, possible arguments may be as follows.
One point is that the relevant scale $\mu$ for $E_{\rm tot}(r)$
is not equal to $1/r$ but considerably larger.
Another point that may be worth noting is that the system under consideration
would be optimally suited for perturbative
QCD computations.
It is a colour singlet system having a localised spatial extent,
so that we may expect the decoupling of IR degrees of freedom
to be realised in a most natural way.\footnote{
It may be contrasted with e.g.\ perturbative QCD calculations of 
parton scattering amplitudes, where spatially separated 
coloured partons exist as asymptotic states.
}
\section*{Appendix} \label{appendix}

In this paper we expand the perturbative series of the QCD potential
and the quark pole mass in the strong coupling constant defined 
in the $\overline{\rm MS}$ scheme.
The coupling constant obeys the renormalisation--group equation:
\begin{eqnarray}
\label{rge}
\mu^2\,\frac{d}{d\mu^2}\,{\alpha_s(\mu)}
= \beta (\alpha_S(\mu)) =
- \, \alpha_S(\mu) \sum_{n=0}^{\infty} \beta_n  
\left( \frac{\alpha_S(\mu)}{4\pi} \right)^{n+1} .
\label{RGeq}
\end{eqnarray}
We include the coefficients of the beta function up to 3 loops
in our analysis\footnote{
Although the 4--loop coefficient is available, we consider the
use of the 3--loop beta function to be more consistent with the analysis
% you do not consider that A is B. You consider A to be B.
of the fixed--order perturbative series up to ${\cal O}(\alpha_S^3)$.
},
i.e.\
$\beta_0 = 11$, $\beta_1 = 102$, $\beta_2 =2857/2$ ($n_l=0$), and
$\beta_n =0$ for $n \geq 3$.

In rewriting $\alpha_S(\overline{m})$ in terms of 
$\alpha_S(\mu)$ in the fixed--order 
expression of $E_{\rm tot}(r)$, we use the perturbative
solution of Eq.~(\ref{RGeq}):
\begin{eqnarray}
\alpha_S(\overline{m})=\alpha_S(\mu)
\Biggl[ \,
1 + \frac{\alpha_S(\mu)}{\pi}\, \frac{\beta_0}{2} \, 
\log \Bigl( \frac{\mu}{\overline{m}} \Bigr)
%\nonumber \\
+ \biggl( \frac{\alpha_S(\mu)}{\pi} \biggr)^2
\biggl\{ 
\frac{\beta_0^2}{4} \,\log^2 \Bigl( \frac{\mu}{\overline{m}} \Bigr)
+ \frac{\beta_1}{8} \,\log \Bigl( \frac{\mu}{\overline{m}} \Bigr)
\biggr\} + \cdots
\Biggr] .
\label{pertrel}
\end{eqnarray}
This relation is inserted to Eq.~(\ref{massrel}) and the
series expansion is truncated at order $\alpha_S(\mu)^3$.
Then $E_{\rm tot}(r)$ is given as a function of $r$, $\overline{m}$, $\mu$
and $\alpha_S(\mu)$.

The value of $\alpha_S(\mu)$ is determined by the renormalisation--group 
evolution from the input Lambda parameter 
defined in the $\overline{\rm MS}$ scheme,
$\Lambda_{\overline{\rm MS}}$;
see e.g.\ ref.~\cite{Chetyrkin:1997sg} for the definition of
$\Lambda_{\overline{\rm MS}}$.
The renormalisation--group evolution of $\alpha_S(\mu)$ is
calculated in two different ways in this paper.
When we refer to ``numerical running'', we solve the
renormalisation--group equation (\ref{RGeq}) numerically.
In this case, 
$\Lambda_{\overline{\rm MS}}=0.65573... \times \mu_{\rm Landau}$
($n_l=0$),
where $\mu_{\rm Landau}$ is the position of the infrared singularity 
(Landau singularity)
of the running coupling constant $\alpha_S(\mu)$.
On the other hand, when we refer to ``analytical running'',
we use an approximate analytic solution of the 
renormalisation--group equation:
\bea
\frac{\alpha_S(\mu)}{\pi} \approx
\frac{4}{\beta_0\,L}-\frac{\beta_1\,\log L}{(\beta_0\,L)^2}
+\frac{1}{(\beta_0\,L)^3}\, \biggl[ \,
\frac{\beta_1^2}{4}\, ( \log^2 L - \log L - 1 ) + \beta_2 \, \biggr] ,
\eea
where $L = \log (\mu^2/\Lambda_{\overline{\rm MS}}^2)$ and terms of 
${\cal O}(1/L^4)$ have been neglected.

%%%%%%%%%%%%%%%%%%%%%%%%%%%%
\section*{Acknowledgements}

We are grateful to T. Kaneko, S.~Necco and H.~Suganuma for providing the
lattice data and to A.~Pineda and S.~Necco for discussions.
S.R.\ was supported by the Japan Society for the Promotion of Science
(JSPS).


\begin{thebibliography}{99}

\bibitem{pinedasoto}
  A. Pineda and J. Soto, 
  Nucl.~Phys.~Proc.~Suppl.~{\bf 64}, 428 (1998).

\bibitem{bpsv}
  N. Brambilla, A. Pineda, J. Soto and A. Vairo, 
  Nucl.~Phys. {\bf B566}, 275 (2000).

\bibitem{nnnlo-H}
B. A. Kniehl, A. A. Penin, V. A. Smirnov and M. Steinhauser,\\
 Nucl. Phys. {\bf B635}, 357 (2002).

\bibitem{renormalon1}
  A. Hoang, M. Smith, T. Stelzer and S. Willenbrock, 
  Phys. Rev. {\bf D59}, 114014 (1999).

\bibitem{renormalon2}
  M. Beneke, {Phys. Lett.} {\bf B434}, 115 (1998).  

\bibitem{bsv1}
 N. Brambilla, Y. Sumino and A. Vairo, 
 Phys.~Lett. {\bf B513}, 381 (2001).

%\cite{Sumino:2001eh}
\bibitem{Sumino:2001eh}
Y.~Sumino,
%``A connection between the perturbative QCD potential and  phenomenological potentials,''
Phys.\ Rev.\ D {\bf 65}, 054003 (2002).
%[arXiv:hep-ph/0104259].
%%CITATION = HEP-PH 0104259;%%

\bibitem{necco-sommer}
  S.~Necco and R.~Sommer,
  {Phys. Lett.} {\bf B523}, 135 (2001).  

\bibitem{bsv2} 
  N. Brambilla, Y. Sumino and A. Vairo, 
  Phys. Rev. {\bf D65}, 034001 (2002).

%\cite{Recksiegel:2001xq}
\bibitem{Recksiegel:2001xq}
S.~Recksiegel and Y.~Sumino,
%``Perturbative QCD potential, renormalon cancellation and  phenomenological potentials,''
Phys.\ Rev.\ D {\bf 65}, 054018 (2002).
%[arXiv:hep-ph/0109122].
%%CITATION = HEP-PH 0109122;%%

%\cite{Pineda:2002se}
\bibitem{Pineda:2002se}
A.~Pineda,
%``The static potential: Lattice versus perturbation theory in a  renormalon-based approach,''
J.\ Phys.\ G {\bf 29}, 371 (2003).
%[arXiv:hep-ph/0208031].
%%CITATION = HEP-PH 0208031;%%

%\cite{Lee:2002sn}
\bibitem{Lee:2002sn}
T.~Lee,
%``Surviving the renormalon in heavy quark potential,''
Phys.\ Rev.\ D {\bf 67}, 014020 (2003).
%[arXiv:hep-ph/0210032].
%%CITATION = HEP-PH 0210032;%%

\bibitem{impQCD}
%\cite{Recksiegel:2002za}
%\bibitem{Recksiegel:2002za}
S.~Recksiegel and Y.~Sumino,
%``Improved perturbative QCD prediction of the bottomonium spectrum,''
Phys.\ Rev.\ D {\bf 67}, 014004 (2003).
%[arXiv:hep-ph/0207005].
%%CITATION = HEP-PH 0207005;%%

\bibitem{Brambilla}
N.~Brambilla, hep-ph/0012211. 

\bibitem{Bali}
G.~Bali,
Phys.~Rept.~{\bf 343}, 1 (2001).

\bibitem{Poincare}
N.~Brambilla,, D.~Gromes and A.~Vairo,
Phys.~Rev.~{\bf D64}, 076010 (2001).

\bibitem{al}
  U. Aglietti and Z. Ligeti, Phys. Lett. {\bf B364}, 75 (1995).

\bibitem{hyperfine}
S.~Recksiegel and Y.~Sumino,
hep-ph/0305178.

%\cite{Capitani:1998mq}
\bibitem{Capitani:1998mq}
S.~Capitani, M.~L\"uscher, R.~Sommer and H.~Wittig  [ALPHA Collaboration],
%``Non-perturbative quark mass renormalization in quenched lattice QCD,''
Nucl.\ Phys.\ B {\bf 544}, 669 (1999); Erratum ibid.\ {\bf 582}, 762 (2000).
%[arXiv:hep-lat/9810063].
%%CITATION = HEP-LAT 9810063;%%

\bibitem{ps}
  M.~Peter, Phys. Rev. Lett.~{\bf 78}, 602 (1997); 
  Nucl. Phys. {\bf B501} 471 (1997);
  Y.~Schr\"oder, {Phys.~Lett.}~{\bf B447}, 321 (1999).  

\bibitem{mr}
  K. Melnikov and T.~v.~Ritbergen, Phys. Lett. {\bf B482}, 99 (2000).

\bibitem{beneke}
  M.~Beneke, Phys.~Rept.~{\bf 317}, 1 (1999).

\bibitem{Sumino-pro}
Y. Sumino, hep-ph/0004087. 

%\cite{Bali:1992ru}
\bibitem{Bali:1992ru}
G.~S.~Bali and K.~Schilling,
%``Running coupling and the Lambda parameter from SU(3) lattice simulations,''
Phys.\ Rev.\ D {\bf 47}, 661 (1993).
%[arXiv:hep-lat/9208028].
%%CITATION = HEP-LAT 9208028;%%

\bibitem{suganuma}
T.T. Takahashi et al., 
Phys. Rev. {\bf D65}, 114509 (2002).

\bibitem{kaneko}
JLQCD Collaboration, S.~Aoki, et al., {\it to appear in hep-lat}.

\bibitem{necco-sommer-data}
S.~Necco and R.~Sommer,
%``The N(f) = 0 heavy quark potential from short to intermediate  distances,''
Nucl.\ Phys.\ B {\bf 622}, 328 (2002).
%[arXiv:hep-lat/0108008].
%%CITATION = HEP-LAT 0108008;%%

%\cite{Chetyrkin:1997sg}
\bibitem{Chetyrkin:1997sg}
K.~G.~Chetyrkin, B.~A.~Kniehl and M.~Steinhauser,
%``Strong coupling constant with flavour thresholds at four loops in the  MS-bar scheme,''
Phys.\ Rev.\ Lett.\  {\bf 79}, 2184 (1997).
%[arXiv:hep-ph/9706430].
%%CITATION = HEP-PH 9706430;%%


\end{thebibliography}
\end{document}